\documentclass[twocolumn,epjc3,,nopacs]{svjour3}  
\smartqed  
\RequirePackage{graphicx}
\RequirePackage{latexsym,amsmath,amssymb,bm,mathrsfs,braket}
\RequirePackage{multirow,graphicx,longtable,booktabs}
\RequirePackage[english]{babel}
\RequirePackage{xcolor}
\RequirePackage{fix-cm}
\RequirePackage{mathptmx}
\RequirePackage[T1]{fontenc}
\RequirePackage{epstopdf}
\RequirePackage{soul,bookmark}
\RequirePackage{cite}
\journalname{Eur. Phys. J. A}
\begin{document}
	
\title{Fission fragment distributions within time-dependent density functional theory}

\author{
	Yun Huang\thanksref{addr1} \and Xiang-Xiang Sun\thanksref{addr1,addr2} \and Lu Guo\thanksref{addr1,addr3,e1}}
\thankstext{e1}{e-mail: luguo@ucas.ac.cn (corresponding author)}

\institute{
	School of Nuclear Science and Technology, University of Chinese Academy of Sciences, Beijing 100049, China \label{addr1}
	\and 
	Institut f{\"u}r Kernphysik, Institute for Advanced Simulation and J{\"u}lich Center for Hadron Physics, Forschungszentrum J{\"u}lich, D-52425 J{\"u}lich, Germany \label{addr2}
	\and
	Institute of Theoretical Physics, Chinese Academy of Sciences, Beijing 100190, China \label{addr3}
}
%
\date{the date of receipt and acceptance should be inserted later}

\maketitle
		
\begin{abstract}
			A notable issue,
the proper description of mass and charge distributions of fission fragments within non-adiabatic descriptions of fission dynamics,
is investigated by performing double particle number projection on the outcomes of time-dependent Hartree-Fock (TDHF) simulation.
The induced fission process of the benchmark nucleus $^{240}\mathrm{Pu}$ is studied.
In the three-dimensional Cartesian coordinate without any symmetry restrictions,
we get the static fission pathway from the two-dimensional potential energy surface,
and then the fission dynamics from saddle to scission point are obtained using TDHF.
We show that the charge numbers of primary heavy fragments from TDHF simulation strongly depend on the deformations of initial configurations via
the two asymmetric fission channels, which can be distinguished according to the dynamical fission trajectories.
The charge distribution of fission fragments is well reproduced using the double particle number projection technique.
After applying the Gaussian kernel estimation based on the distribution from 
the double particle number projection technique, the mass distribution is also 
consistent with the experimental results.
Besides, the results of the total kinetic energy of fission fragments are reasonably consistent with the experiments.
\end{abstract}
\section{Introduction}
\label{intro}

Nuclear fission is the process of a heavy atomic nucleus
splitting into two or three fragments
with releasing a large amount of energy.
Describing the entirety of the fission process within a unified framework is challenging due to its time-dependent nature and the presence of multiple distinct phases.
\cite{Krappe2012,Schunck2016_RPP79-116301,Simenel2018_PPNP103-19,Bender2020_JPG47-113002,Schunck2022_PPNP125-103963}.
Among these phases, the transition between the mother nucleus and the fragments is the most striking because many effects, 
including rapid shape change, neck formation, energy dissipation,
and quantum shells, interplay and determine its complex mechanism.
The characteristics of fission fragments (FFs) and the correlations
between them are interesting both experimentally and
theoretically since they are essential to reveal
the puzzles of fission mechanism.
Besides, fission is also relevant to other important issues,
such as the $r$-process of nucleosynthesis \cite{Erler2012_PRC85-025802}
and the existence of superheavy elements \cite{Hofmann2000_RMP72-733}.

Fission occurs when the system overcomes the fission barrier(s), and
to date, the microscopic theories describe this process usually in an adiabatic way,
i.e., building a static potential energy surface (PES) in the collective space.
The next phase is a process from saddle to scission point with dissipative motion and nowadays can be described microscopically by using the
time-dependent generator coordinate method (TDGCM)
\cite{Schunck2016_RPP79-116301,Verriere2020_FP8-00233}
and time-dependent density functional theory (TDDFT)
\cite{Bulgac2016_PRL116-122504,Nakatsukasa2016_RMP88-045004,
	Simenel2018_PPNP103-19,Scamps2018_Nature564-382,Ren2022_PRL128-172501}.
The former, as an adiabatic time-dependent theory, describes the fission dynamics based on a static multi-dimensional PES and computes the realistic fission-fragment distributions under the Gaussian overlap approximation (GOA),
which is powerful for describing the distributions of FFs
\cite{Schunck2016_RPP79-116301,Regnier2016_PRC93-054611,
	Verriere2019_PRC100-024612,Verriere2021_PRC103-054602}.
The latter includes all collective degrees of freedom and
can simulate the average fission trajectory as a non-adiabatic
dynamics; meanwhile, the role of shell and pairing effects on
fission dynamics can be explored
\cite{Goddard2015_PRC92-054610,Goddard2016_PRC93-014620,
	Bulgac2016_PRL116-122504,Scamps2018_Nature564-382,
	Scamps2015_PRC92-011602,
	Qiang2021_PRC103-L031304,Ren2022_PRL128-172501}.
Recently, the excitation energy sharing mechanism between FFs \cite{Bulgac2016_PRL116-122504,Bulgac2019_PRC100-034615,Bulgac2020_PRC102-044609}
and their intrinsic spins
\cite{Bulgac2021_PRL126-142502,Bulgac2022-PRL128-022501}
are also investigated by TDDFT.
However, the lack of particle number fluctuations leads to TDDFT not giving a good
description of distributions of fission yields
\cite{Lacroix2014_EPJA50-95,Scamps2018_Nature564-382,Bulgac2019_PRC100-034615}.
To resolve this problem, the particle number projection (PNP) technique
\cite{Simenel2010_PRL105-192701} has been
applied to get the distributions of FFs
\cite{Scamps2015_PRC92-011602,Bulgac2019_PRC100-034612,Bulgac2021_PRC104-054601}.
As a result,
the proton and neutron distributions for each fission trajectory can be calculated, but a quantitative description of the measurements, particularly for the width, is not achieved.
Even the combination of TDDFT and TDGCM is also not valid for a proper description of fission-fragment distributions \cite{Ren2022_PRC105-044313}.
This means that based on a non-adiabatic description of fission dynamics, the fission-fragment properties are still not well understood.

$^{240}$Pu is a benchmark nucleus whose fission barriers,
half-life, and fission-fragments are well known
\cite{Wagemans1984_PRC30-218,Schillebeeckx1992_NPA545-623,Nishio1995_JNST32-404,Holden2000_PAC72-1525,
	Tsuchiya2000_JNST37-941,Capote2009_NDS110-3107,JENDL2011}
such that it has been studied by many different approaches
\cite{Rodriguez-Guzman2014_EPJA50-142,
	Goddard2015_PRC92-054610,Goddard2016_PRC93-014620,
	Younes2012,Regnier2016_PRC93-054611,
	Bulgac2016_PRL116-122504,Scamps2018_Nature564-382,
	Bulgac2019_PRC100-034615,Zhao2020_PRC101-064605,
	Qiang2021_PRC103-L031304,Verriere2021_PRC103-054602,
	Bulgac2021_PRL126-142502,Bulgac2022-PRL128-022501,
	Ren2022_PRC105-044313,Tong2022_PRC106-044611,
	Marevic2020_PRL125-102504,Marevic2021_PRC104-021601} in recent years.
Among these works,
the TDDFT theory provides a good description of the induced fission
dynamics of $^{240}\mathrm{Pu}$ and can study many properties of FFs
such as the total kinetic energy (TKE),
collective excitation modes and intrinsic spin.
However, an accurate prediction of the fragments distribution
remains a challenge.
In this work, aiming to study the fission mechanism
within the framework of TDDFT and explore the mass and charge distributions of FFs,
we have implemented the double PNP technique
\cite{Scamps2013_PRC87-014605,Scamps2015_PRC92-011602,Verriere2019_PRC100-024612}
based on the time-dependent Hartree-Fock theory (TDHF).
Consequently, our calculations show a nice agreement with the
measured mass and charge distributions of $^{240}$Pu,
thus proposing a treatment for predicting the FFs' distribution
within the framework of TDDFT.
This paper is organized as follows.
The theoretical formulas are introduced in Sec.~\ref{theory}.
The numerical details, static fission pathway in two-dimensional (2D) PES, the shape evolution, the mass and charge distributions, and the TKE of FFs are presented in Sec.~\ref{results}.
Finally, we summarize this work in Sec.~\ref{summary}.

\section{Theoretical framework}\label{theory}
	The ground state properties of $^{240}\mathrm{Pu}$ are calculated using
the Skyrme Hartree-Fock (HF) theory with the BCS approximation (HF+BCS) \cite{Chabanat1998_NPA635-231}.
For calculating the PES,
the augmented Lagrangian method (ALM) \cite{Staszczak2010_TEPJA46-85}
has been implemented based on the HF$+$BCS in three-dimensional coordinate space and the Routhian reads
\begin{equation}\label{PNP1}
	E^{\prime}=E_{\text{HF}}+
	\sum_{i}
	\left[\lambda_{i}
	\left(\langle\hat{Q}_{i0}\rangle-Q_{i0}\right)+
	\frac{1}{2}C_{i}\left(\langle\hat{Q}_{i0}\rangle-Q_{i0}\right)^{2}
	\right],
\end{equation}
where $i = 2,3$ denotes the order of mass multipole moment,
$\lambda_{i}$ is Lagrangian multiplier,
and $C_{i}$ is the penalty parameter.
The definitions of mass quadrupole and octupole moment operators are
\begin{equation}
	\begin{split}
		\hat{Q}_{20}&=\sqrt{\frac{5}{16\pi}}(2\hat{z}^{2}-\hat{x}^{2}-\hat{y}^{2}),\\
		\hat{Q}_{30}&=\sqrt{\frac{7}{16\pi}}\hat{z}(2\hat{z}^{2}-3\hat{x}^{2}-3\hat{y}^{2}).\\
	\end{split}
\end{equation}	
The constraints on $\hat{Q}_{20}$ and $\hat{Q}_{30}$ are performed simultaneously
to get the static PES.

The fission dynamics is described using the TDHF method,
in which the many-body wave function is approximated as a single
time-dependent Slater determinant composed of single particle states $\psi_{\alpha}(\bm{r}s \tau,t)$ with spatial coordinate $\bm{r}$, spin $s$ and isospin $\tau$.
The TDHF equations read
\begin{equation}
	i\hbar\frac{\partial}{\partial t}\psi_{\alpha}(\bm{r}s \tau, t)=\hat{h}\psi_{\alpha}(\bm{r}s \tau, t),
\end{equation}
where $\hat{h}$ is the single-particle Hamiltonian.
The pairing correlations in the dynamical evolution are considered
with the frozen occupation approximation (FOA) \cite{Scamps2013_PRC87-014605}. 
Although the FOA is undoubtedly the most straightforward way to consider dynamic pairing correlations which are crucial to investigating fission dynamics \cite{Scamps2015_PRC92-011602,Bulgac2016_PRL116-122504}, 
it is still promising for providing theoretical guidance \cite{Simenel2014_PRC89-031601,Goddard2015_PRC92-054610,Goddard2016_PRC93-014620,Simenel2018_PPNP103-19}.
In the TDHF+FOA simulation of the fission process, the shape of the mother nucleus rapidly changes and the scission configuration can automatically occur.
Most importantly, the properties of FFs,
such as charge, mass, and TKE, are
natural outputs of TDHF+FOA calculations.

The TDHF+FOA approach describes the trajectories of FFs in a classical way and
the quantum fluctuations are not included, thus
it only gives the mean values and cannot give the distributions of FFs properly.
The PNP technique based on TDHF
\cite{Simenel2010_PRL105-192701}
can treat the particle number fluctuations and its application on fission
\cite{Scamps2013_PRC87-014605}
can be used to study the FFs distributions.
In PNP technique, the projection operator for $N$ particles
with isospin $\tau$ in the subspace $\mathrm{V}$
is
\begin{equation}\label{1}
	\hat{P}^{\tau}_{\mathrm{V}}(N)
	\equiv\frac{1}{2\pi}\int_{0}^{2\pi}d\theta~ e^{i\theta( \hat{N}^{\tau}_{\mathrm{V}}-N)},
\end{equation}
where $\hat{N}^{\tau}_{\mathrm{V}}$ is the particle number operator and $\theta$ is the gauge angle. In Eq.~(\ref{1}), only the particle number fluctuation of the fragment in the subspace is considered, without accounting for the influence attributed to the superfluidity of the mother nucleus.
To consider this point,
the double PNP technique is utilized to get the probability of
$N$ protons or neutrons in the subspace for a given mother nucleus in the whole space.
The probability is defined as
\begin{equation}\label{2}
	P^{\tau,~\mathrm{Double}}_{\mathrm{V}}(N)=
	\frac{\bra{\Psi}\hat{P}^{\tau}_{\mathrm{V}}(N)\hat{P}^{\tau}(N_{\mathrm{tot}})\ket{\Psi}}{\bra{\Psi}\hat{P}^{\tau}(N_{\mathrm{tot}})\ket{\Psi}},
\end{equation}
where $N_{\mathrm{tot}}$ is the number of neutrons $(\tau=-1)$ or protons $(\tau=1)$ of the mother nucleus and $|\Psi\rangle$ is the final state in the TDHF calculation.
The overlap in Eq.~(\ref{2}) is calculated by using the Pfaffian method \cite{Robledo2009_PRC79-021302,GonzalezBallestero2011_CPC182-2213}, which can efficiently solve the sign ambiguity problem of the Onishi formula \cite{Onishi1966_NP80-367}.
Then, we use the Fomenko's method \cite{Fomenko1970_JPA3-8} with 31 mesh points for the integral over the guage angles.

For comparison with experimental data, we should fold the mass and charge distributions of all selected fission events. The charge and mass yields are given by
\begin{equation}
	\begin{split}
		Y_{\mathrm{V}}(Z) &=\sum_{S}w(S)\times P^{\tau=1,~\mathrm{Double}}_{S,\mathrm{V}}(Z),\\
		Y_{\mathrm{V}}(A) &=\sum_{S}w(S)\times P^{\mathrm{Double}}_{S,\mathrm{V}}(A),\\
	\end{split}
\end{equation}
where $w(S)$ denotes the weight of a given fission trajectory $S$.	 
The $Y_{\mathrm{V}}(Z)$ and $Y_{\mathrm{V}}(A)$ are normalized to $2$.

\section{Results and discussions}\label{results}
The constrained HF with the BCS approximation for pairing correlations (CHF+BCS)
and TDHF calculations with FOA are performed
using a modified version of the code \texttt{Sky3D}
\cite{Guo2007_PRC76-014601,Guo2008_PRC77-041301,Maruhn2014_CPC185-2195},
in which the time-even, time-odd, and tensor terms in Hamiltonian have been incorporated
\cite{Dai2014_SciChinaPMA57-1618,Guo2018_PLB782-401,Li2022_PLB833-137349,Sun2022_CTP74-097302}
and the techniques introduced in Sec.~\ref{theory} have also been implemented.
More details of our code and its applications can be found in
Refs. \cite{Dai2014_PRC90-044609,
	Yu2017_SciChinaPMA60-092011,
	Guo2018_PLB782-401,
	Guo2018_PRC98-064609,
	Guo2018_PRC98-064607,
	Li2019_SCPMA62-122011,
	Wu2019_PRC100-014612,
	Godbey2019_PRC100-054612,
	Wu2020_SCPMA63-242021,
	Wu2022_PLB825-136886,
	Sun2022_PRC105-034601,
	Sun2022_PRC105-054610,
	Sun2023_PRC107-L011601,
	Sun2023_PRC107-064609}.
In this work, we use the density functional SLy5,
which has been widely used in the study of nuclear dynamics
\cite{Guo2018_PRC98-064609,
	Guo2018_PRC98-064607,
	Guo2018_PLB782-401,
	Godbey2019_PRC100-054612,
	Wu2022_PLB825-136886,
	Sun2022_PRC105-034601,
	Sun2022_PRC105-054610,
	Li2022_PLB833-137349}.
Unlike the commonly used SkMs and SLy4d, this set of interactions is a new attempt in fission.
Besides, the comparison of results in the present work with SLy5 with subsequent investigations with SLy5t will reveal the influence of tensor forces on fission.
In CHF+BCS calculations,
the three-dimensional grid $36\times32\times36$ $\mathrm{fm}^{3}$
with the grid spacing of 1 fm in each direction is adopted
and the grid $40\times40\times40$ $\mathrm{fm}^{3}$ is used
when $Q_{20}\geq73 ~\mathrm{b}$
to keep the accuracy.
In static calculations,
we adopt the volume delta pairing force
with pairing strengths $V^{n}_{0}=288.523$ $\mathrm{MeV}$ $\mathrm{fm}^{3}$
and $V^{p}_{0}=298.760$ $\mathrm{MeV}$ $\mathrm{fm}^{3}$ for neutrons and protons
\cite{Maruhn2014_CPC185-2195}, respectively.
The TDHF calculations are performed in a box with the size of
$48\times48\times160$ $\mathrm{fm}^{3}$.
The time step is $0.25$ $\mathrm{fm/c}$ and the time propagator
is evaluated by the Taylor series expansion with the order up to 8.
The center-of-mass correction is not taken into account in both static and dynamic calculations.
These numerical conditions have been checked for
achieving good convergence for all computations.

\begin{figure}[htbp]
	\includegraphics[width=0.48\textwidth]{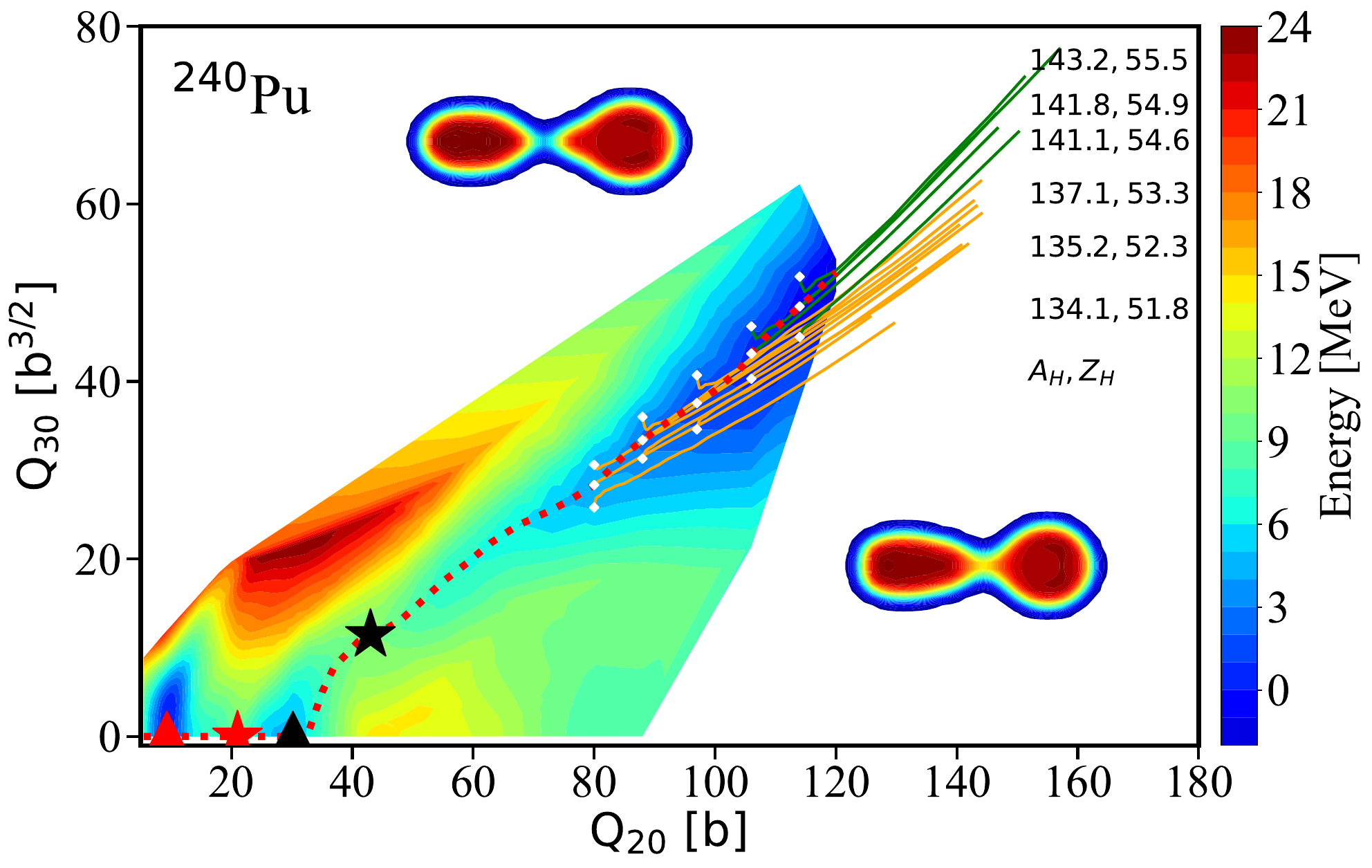}
	\caption{Potential energy surface of $^{240}\mathrm{Pu}$
		in the $(Q_{20},~Q_{30})$ plane calculated with SLy5 and
		the TDHF fission trajectories for various initial configurations.
		The red dotted line shows the static fission pathway.
		The ground state, inner barrier, isomeric state,
		and outer barrier are denoted by red triangle,
		red star, black triangle and black star, respectively.
		Selected initial states of TDHF calculations are marked by the white diamonds. The dynamical fission trajectories are presented by the yellow and green solid lines corresponding to two asymmetric fission channels, respectively, and their representative density distribution profiles at the scission points are also shown.
		The mass and charge numbers of heavy fragments are shown on the right.}
	\label{Fig:PES}
\end{figure}

By carrying out CHF+BCS calculations,
we have constructed the PES of $^{240}\mathrm{Pu}$ in $(Q_{20},~Q_{30})$ plane
on a three-dimensional coordinate space without any symmetry restrictions and other shape degrees of freedom are incorporated automatically
in the self-consistent calculations.
The obtained 2D PES is presented in Fig.~\ref{Fig:PES},
where the energies
are given relative to the ground state energy.
The ground
and isomeric states
are labeled by the red and black triangles, respectively.
As shown in Fig. \ref{Fig:PES}, the static fission pathway (red dotted line) has a double-humped structure,  and the inner and outer barriers are located at
$Q_{20}\approx21~\mathrm{b}$ and $Q_{20}\approx43~\mathrm{b}$.
In our calculation, the height of the primary barrier is about 10 MeV,
which is larger than the HF calculation with SkM* parametrization
\cite{Goddard2015_PRC92-054610} and the empirical values given in
Ref. \cite{Capote2009_NDS110-3107,Bertsch2015_JPG42-077001}.
In fact, for most Skyrme forces,
the calculated barrier heights are higher than the empirical values,
perhaps as a result of the adiabatic calculation lacking dynamical effects.
In comparison to the SkM* force, the fitting procedure of SLy5
does not constrain any surface properties,
despite the barrier height lowers with the decrease of
the surface energy coefficient \cite{Jodon2016_PRC94-024335}.
Moreover, the study of the deformation-induced fission dynamics with TDHF
usually starts beyond the outer barrier,
meaning that the
dynamical process is not as sensitive to the barrier height as the spontaneous fission.
Therefore, the fission barrier height of $^{240}$Pu
with SLy5 is acceptable.
Additionally, other shape degrees of freedom
also affect the fission barrier height \cite{Lu2012_PRC85-011301,Lu2014_PRC89-014323}
and the insight into the PES requires huge computational cost,
which are beyond the scope of present work.
Generally speaking, the main features of PES
with SLy5 are similar to
the results given by other theoretical calculations
\cite{Schunck2014_PRC90-054305,Lu2014_PRC89-014323,
	Bulgac2019_PRC100-034615,Qiang2021_PRC103-L031304}.
For the asymmetric fission pathway,
the shape of outer barrier is sharp,
reflecting rapid changes of the single particle levels.
The fission valley gradually appears beyond the outer barrier
and contains the starting configurations for the TDHF+FOA simulation.

A 2D PES in the $(Q_{20},~Q_{30})$ plane
not only contains
more details about the static fission pathway,
but also naturally has more complex configurations
due to the availability of different $Q_{30}$.
In TDHF description of fission process \cite{Goddard2015_PRC92-054610},
fission occurs
when the deformations of initial states are larger than
a dynamical fission threshold ($Q_{20}\approx80\ \mathrm{b}$ of $^{240}\mathrm{Pu}$ in this work).
Additionally, the threshold anomaly also occur in the $Q_{30}$ direction, thus leading to more non-fission events.
In this work we select $15$ deformation
configurations from the 2D PES, labeled by white diamonds in Fig. \ref{Fig:PES}
as the initial states of TDHF+FOA simulations.
Specifically, along the static asymmetric fission pathway (red dotted line),
five points in the $Q_{20}$ direction
and three in the $Q_{30}$ direction with almost equal spacing,
i.e., 15 points are chosen.
This set of initial configurations corresponds to a mean excitation energy of 2.1 MeV with a variance of 3 MeV for $^{240}$Pu.
Such a broad excitation energy range and variance may cover various initial configurations such that the fission dynamics occur in different fission channels. 

In our TDHF calculations, fission happens for all of these 15 configurations
and the dynamical fission trajectories are represented by solid lines ending at the scission point in Fig. \ref{Fig:PES}, which is defined as the point when the minimum density between the fragments along the $z$ axis is less than 0.03 $\mathrm{fm}^{-3}$.
It is found that
during the dynamical evolution, the reflection-asymmetric shape
increases with the elongation of the fissioning nucleus. 
The dynamical pathways can be divided into two groups characterized by the size of $(Q_{20}, Q_{30})$  deformations at the scission points, denoted by yellow and green solid lines in Fig. \ref{Fig:PES}, respectively.

\begin{figure}
	\includegraphics[width=0.49\textwidth]{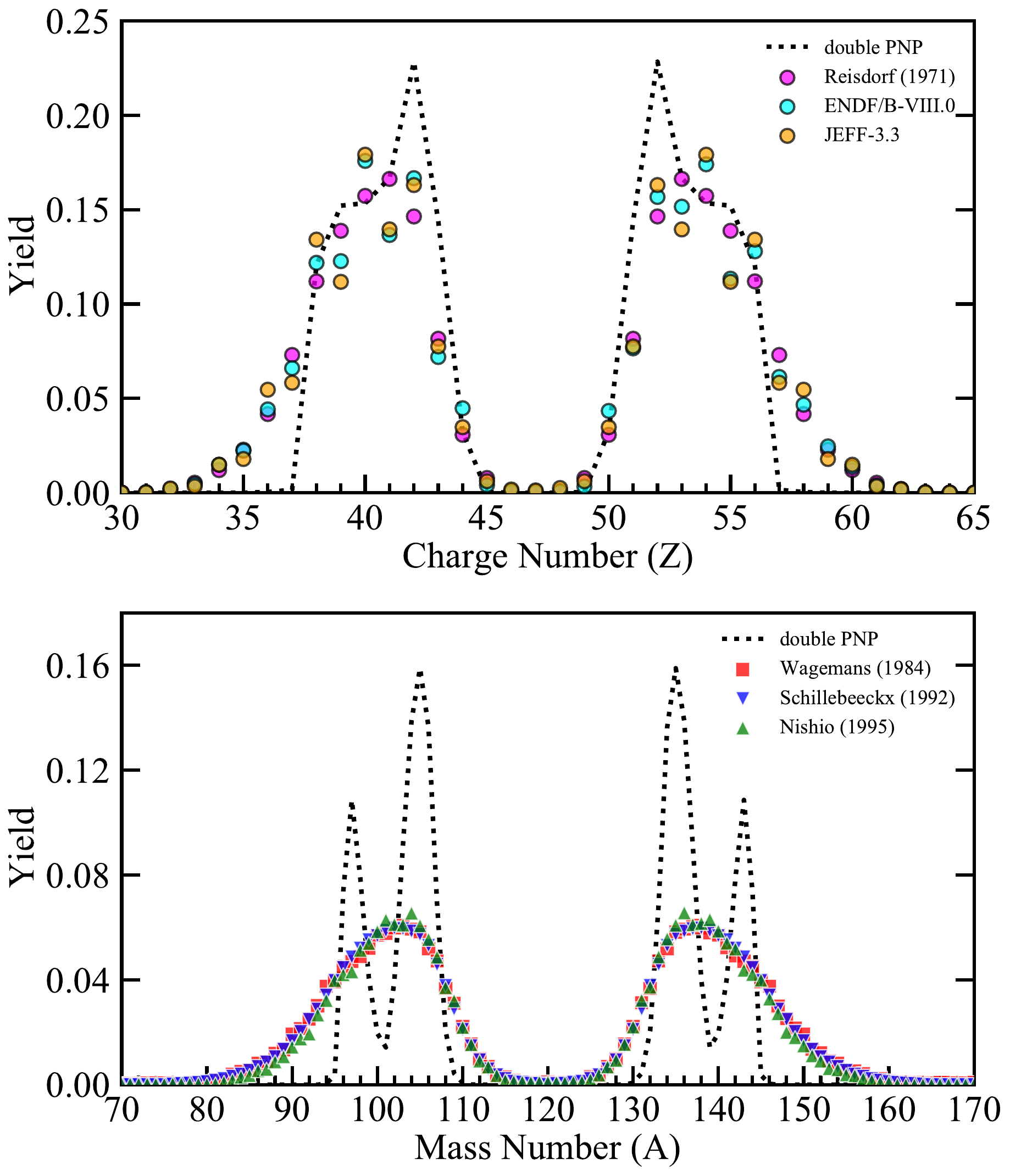}
	\caption{The distribution of charge (upper panel) and mass (bottom panel) for the deformation-induced fission of $^{240}\mathrm{Pu}$.  The black dashed lines represent the results from TDHF+FOA with double PNP calculations. The experimental data taken from Wagemans 1984 \cite{Wagemans1984_PRC30-218}, Schillebeeckx 1992 \cite{Schillebeeckx1992_NPA545-623}, Nishio 1995 \cite{Nishio1995_JNST32-404}, Reisdorf 1971 \cite{Reisdorf1971_NPA177-337}, ENDF/B-VIII.0 library \cite{Brown2018_NDS148-1} and JEFF-3.3 library \cite{Plompen2020_EPJA56-181} are shown for comparison.}
	\label{Fig:rawYield}
\end{figure}

When the separation distance between the centers of mass of two primary fragments is larger than 30 fm, the TDHF+FOA calculation ends and the mass of primary fragments can be directly obtained by integrating one-body local densities in the subspace.
We also give the mass and charge numbers of heavy fragments for selected trajectories in Fig. \ref{Fig:PES}.
The mass and charge numbers of heavy fragments
are around $A_H=134$~--~$143$ and $Z_H=52$~--~56.	
Our calculations show that the mass and charge
numbers of heavy fragments increase with the deformations ($Q_{20},~Q_{30}$) at scission points.
This can be understood as that bigger deformation means larger mass asymmetry,
i.e., a more heavy fragment.
More interestingly, the mass and
charge numbers of heavy fragments of those trajectories with $Q_{30}$ of initial
states smaller than $40\ \mathrm{b}^{3/2}$ (yellow lines) are
centered around 135 and 52 and those with
$Q_{30}$ of initial
states larger than $40\ \mathrm{b}^{3/2}$ (green lines)
are around 143 and 55, clearly indicating the dependence of
primary FFs on initial configurations
in TDHF+FOA simulations.
It has been shown in Ref. \cite{Scamps2018_Nature564-382} that
the shell effects caused by the pear-shaped fragments
lead to most of the heavy fragments formed with
$Z=52$ and 56.
Additionally,
the properties of fragments corresponding to trajectories represented by yellow
and green lines are close to
the standard I (SI) channel with $A_H\approx135$
and TKE$~\approx190$ MeV and the standard II (SII) channel
with $A_H\approx142$ and TKE$~\approx175$ MeV, respectively.
Both two channels are proposed in the Brosa model \cite{Brosa1990_PR197-167}.
Here, the mass number of heavy fragments of SI channel has a mean value of 135.2 with the standard deviation of 0.90 and 142.4 (0.94) for SII channel.
The representative density distribution profiles of the two asymmetric channels are shown in Fig. \ref{Fig:PES}. The one in the lower right with a nearly spherical heavy fragment corresponds to the SI channel, and the other in the upper left shows an evident octupole deformation for the heavy fragment corresponding to the SII channel.
Thus one can deduce that the onset of two channels
is related to the initial deformations and quantum shell
effects in the dynamic process,
thus providing a microscopic interpretation of this phenomenological model.
Similar conclusions and the influence of thermal fluctuations can be found
in Ref. \cite{Qiang2021_PRC103-L031304}.

After exploring the insights of
fission dynamics, we focus on
the mass and charge distributions,
which is well-known experimentally but
a proper description based on the TDDFT calculations
is still not achieved.
To describe the distributions of fragments,
we applied the double PNP [cf. Eq. (\ref{2})]
to the fission outcomes of the TDHF+FOA calculations.
As mentioned above, the initial states selected from the PES follow a nearly equidistant pattern in both $Q_{20}$ and $Q_{30}$ directions, thus the dominant fission events that occurred in the fission valley could be involved.
Considering that each fission trajectory is treated as an independent event in TDHF+FOA simulations \cite{Goddard2015_PRC92-054610,Scamps2018_Nature564-382}, 
all the initial configurations are considered to be equally important.
In addition, the probability distribution from the evolution of an initial Gaussian wave packet by TDGCM under a mean-field PES \cite{Ren2022_PRC105-044313} shows that the trajectories with initial deformations close to the static fission pathway play a dominant role while the remaining ones are less significant.
Therefore, in this work, the probability contributions of mass (or charge) from all 15 dynamical calculations are directly summed together with equal weights.

The mass and charge distributions from TDHF+FOA with double PNP calculations are presented and
compared with experimental data for thermal neutron induced fission of $^{239}\mathrm{Pu}$ in Fig. \ref{Fig:rawYield}.
The distributions are calculated by directly summing up the projected probabilities for all fission trajectories.
The experimental data specific for primary fission fragments and the excitation energy within the range of calculated mean excitation energy of initial configurations are chosen for a direct comparison with our calculations.
In Fig. \ref{Fig:rawYield}, we find that for the charge distributions, 
the width in our calculation is consistent with the data, and the position of the peak only slightly shifts towards the symmetric side.
For the mass distribution, there are two peaks at $A_H = 135$ and $A_H = 142$, 
corresponding to the SI and SII fission channels, respectively. 
Between the two peaks, there is a clear drop, 
indicating a rapid transition from SI to SII. 
Although the mass distribution is not well reproduced, 
a maximum yield for heavy fragments at $A_H = 135$ is consistent with the peak given by the experiments.
Additionally, a notable discrepancy in the distribution width indicates 
that the TDHF+FOA+PNP for fragment distributions still underestimates
the quantum fluctuations on mass numbers during the fission process.
This discrepancy might be understood in two aspects.
On the one hand, due to the threshold anomaly in TDHF+FOA,
the initial configurations
with larger mass asymmetry are non-fission.
This might lead to a narrow mass distribution.
On the other hand, a proper weight reflecting the quantum fluctuation for adiabatic
treatment of collective degrees of freedom for each fission trajectory of TDHF+FOA is still missing.
The probability distributions can be obtained by using semiphenomenological approaches,
solving the dissipative Langevin equations, or TDGCM based on PES.
Along this line, within the framework of TDDFT, it is promising to get a proper description of FFs
distributions based on the combination of TDBCS, double PNP,
and a reasonable probability for fission trajectories, which would be a long-term project.
In recent TDDFT stuides in Refs.
\cite{Ren2022_PRC105-044313,Li2023_FP19-44201},
although the probabilites of TDDFT fission tracjectores are considered with TDGCM,
their results on charge distributions are still not very different from those 
obtained from standard.
In conclusion, the TDHF+FOA+PNP can give a good results for charge distributions but not
for mass distrobutions.

\begin{figure}
	\includegraphics[width=0.49\textwidth]{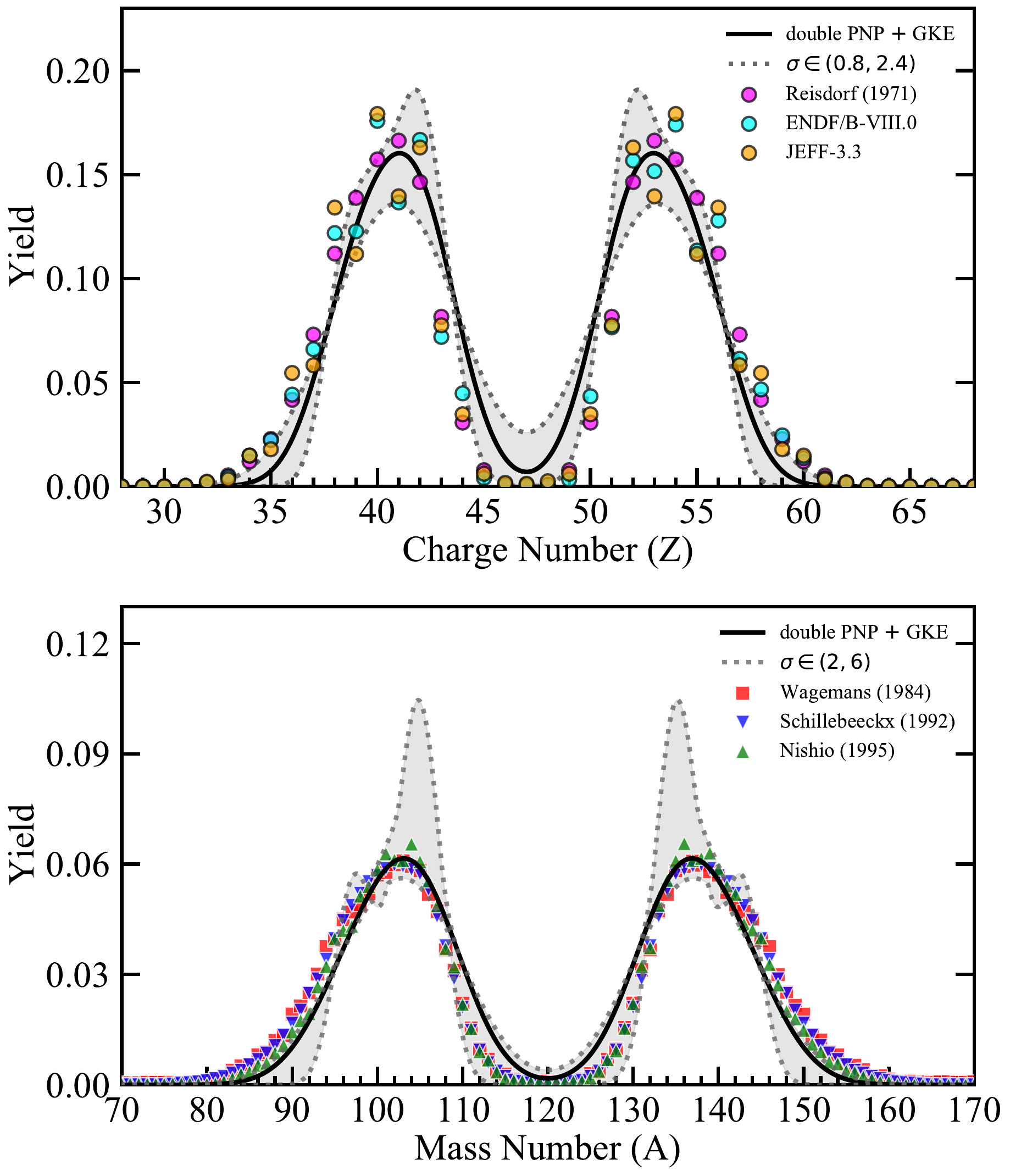}
	\caption{The distribution of charge (upper panel) and mass (bottom panel) for the deformation-induced fission of $^{240}\mathrm{Pu}$. The black lines show the results using the combination of double PNP and Gaussian kernel estimation with $\sigma=5.1$ and 1.6 for mass and charge distributions, respectively.
		The gray area, delineated by gray dotted lines, represents the sensitivity of the results with respect to different values of $\sigma$ in the Gaussian function.
		The same experimental data as Fig. \ref{Fig:rawYield} are shown.}
	\label{Fig:distributions}
\end{figure}

Since the quantum fluctuations in the adiabatic process are not adequately included in present study as discussed above,
leading to the charge and especially mass distributions are not well
reproduced simultaneously,
we hope that the Gaussian kernel estimation (GKE)~\cite{Regnier2016_PRC93-054611,Zhao2020_PRC101-064605} can
partially make up these effects and improve the distributions.
Therefore, in the following, we investigate how GKE influences the distributions from TDHF+FOA+PNP calculations and discuss the choice of the Gaussian width $\sigma$ in GKE.
Figure \ref{Fig:distributions} shows the calculated mass and charge distributions
using GKE based on the distribution from TDHF+FOA+PNP, along with the comparison with experiments.
The results with $\sigma\in[0.8,2.4]$ for charge distribution and $\sigma \in[2,6]$ for mass distribution are displayed in the gray area.
For both distributions, we observe that different $\sigma$ could change the yields of the peaks, but almost do not change their locations. 
As $\sigma$ increases, the yield near the peaks decreases in both cases.	
The results of the charge distribution, compared to the mass distribution, are less affected by the $\sigma$ parameter, due to the relative consistency of charge distribution with the experiment as shown in Fig. \ref{Fig:rawYield}. 
By fitting GKE function with all selected experimental data, we obtain the $\sigma$ of mass and charge distributions to be 5.1 and 1.6, respectively.
In Fig. \ref{Fig:distributions}, taking $\sigma=5.1$ for the mass distribution, it is clear that the corresponding yield (denoted by the black line) well reproduces both the positions of asymmetric peaks and the widths of experiments.
Meanwhile, for the charge distribution with $\sigma=1.6$,
the results from TDHF+FOA are further improved and agree well with the data.
Note that usually the GKE is used to take the experimental  
resolution into account and there might be some
double counting between GKE and PNP.
However, up to date a proper description of fragment distributions using microscopic approaches based on TDDFT is still unavailable. 
Thus, TDHF+FOA+PNP with GKE might be an alternative way.

Note that studies on fission-fragment mass and charge distributions with the TDGCM+GOA method
has reproduced the distributions
\cite{Younes2012,Regnier2016_PRC93-054611,Zhao2020_PRC101-064605,Verriere2021_PRC103-054602},
but in which the fission dynamics is assumed as an adiabatic process.
Recently, the combination of TDDFT and TDGCM has been used to describe the charge distributions of $^{240}$Pu but only the peaks are reproduced \cite{Ren2022_PRC105-044313,Li2023_FP19-44201}.
However, the way to properly account for the weight of initial states in TDDFT is still an open question, which deserves further exploration.
Compared with previous works based on TDDFT
\cite{Goddard2015_PRC92-054610,Goddard2016_PRC93-014620,Bulgac2016_PRL116-122504,
	Scamps2018_Nature564-382,Bulgac2019_PRC100-034615},
our calculations select deformation configurations from both the static
fission pathway and 2D PES as initial states for TDHF simulations,
consider the fluctuations of particle number by using the double PNP,
and use GKE
such that a good consistency between experimental mass and charge fission-fragment distributions and predictions is achieved.

	\begin{figure}
	\includegraphics[width=0.48\textwidth]{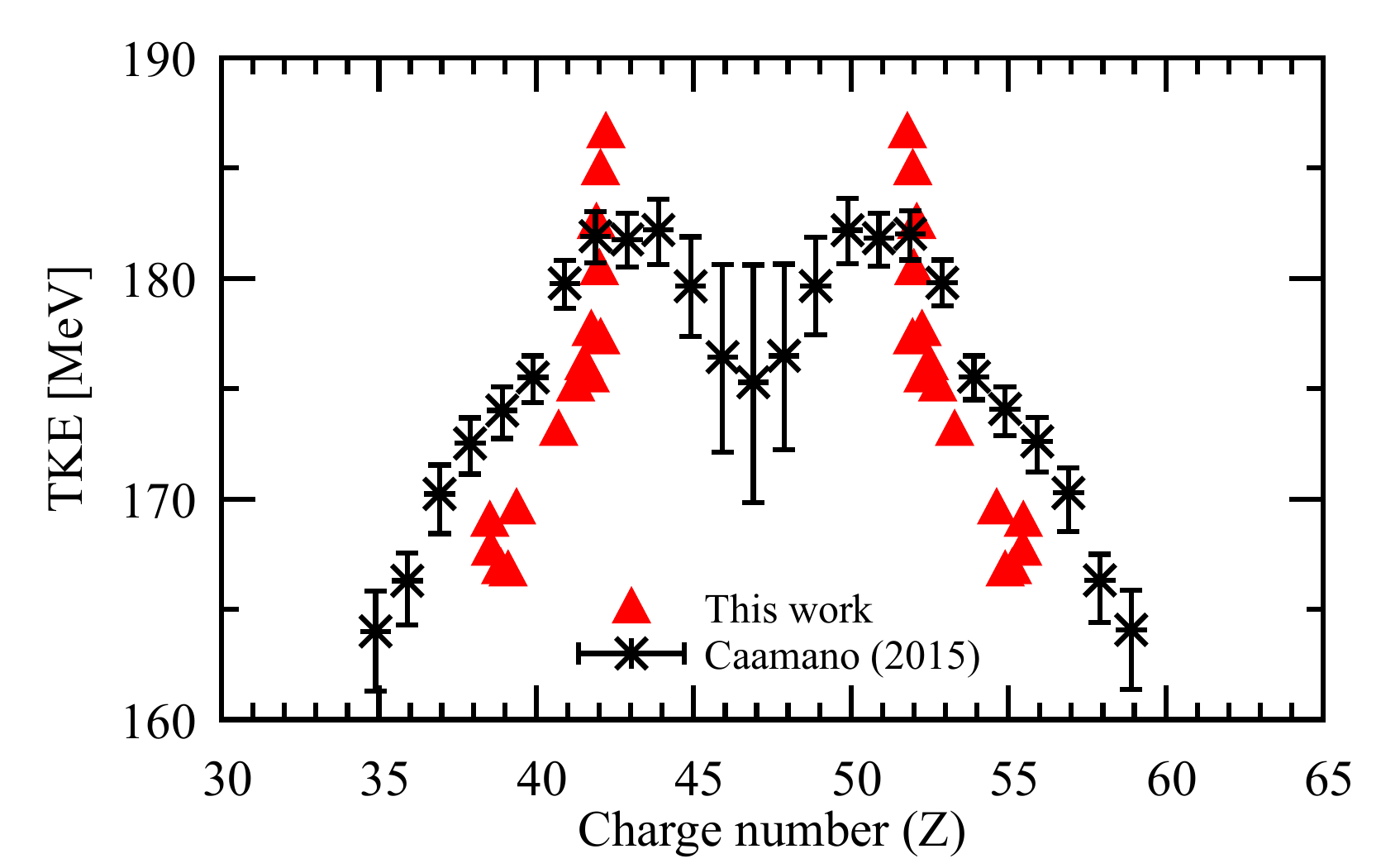}
	\caption{The comparison of calculated TKE (red triangles) with experimental data taken from Caama\~no 2015 \cite{Caamano2015_PRC92-034606}.}
	\label{Fig:Z-TKE-2D-SLy5}
\end{figure}

The TKE is also an important quantity of fission fragments because it directly relates to the occurrence of the scission during the dynamical motion.
Although it has been discussed in many previous studies, here we briefly introduce our results.
In the TDHF theory, TKE is an asymptotic value
by summing the nuclear collective kinetic energy and
the Coulomb energy at a relatively large distance between the fission fragments ($\approx30$~fm in our calculations) \cite{Goddard2015_PRC92-054610}.
In Fig. \ref{Fig:Z-TKE-2D-SLy5}, we display the calculated TKE and
the comparison with experimental data taken from Ref. \cite{Caamano2015_PRC92-034606}.
It is found that
the higher TKE is related to a compact configuration at the scission point (the configuration of SI channel shown in Fig. \ref{Fig:PES})
and more elongation, asymmetric shape leads to a smaller TKE (the configuration of SII channel shown in Fig. \ref{Fig:PES}),
which is similar to the results given in Ref. \cite{Scamps2018_Nature564-382}.
Predicted TKEs marginally overestimate the experimental values when $Z\approx52$ and
there is a slight underestimate of the tail part of the distribution.
In conclusion, the calculated TKE values distribute around the experimental data
and the differences between them are very small (about $10~\mathrm{MeV}$),
indicating a good agreement between theory and experiment.

\section{Summary}\label{summary}
We have presented a microscopic study of the induced fission of $^{240}\mathrm{Pu}$ by using the constrained HF+BCS, TDHF+FOA with double particle number projection, and Gaussian kernel estimation. The 2D PES in the $(Q_{20},~Q_{30})$ plane calculated by using CHF+BCS in a three-dimensional grid shows a double-humped fission pathway and provides the initial configurations beyond the static outer barrier for TDHF simulations. The primary fission fragments are obtained using TDHF with the frozen particle occupation approximation and then the double PNP is performed to get the mass and charge distributions of fission fragments. 
The TDHF+FOA+PNP method can well reproduce the charge distributions but not for mass distributions.
After introducing a GKE, the position of peaks and the width of mass and charge distributions agree rather well with the experiments.
The calculated TKE of fission fragments is also consistent with the experiments.
The present approach can be further improved by treating the dynamical pairing using TDBCS, by which the 
threshold anomaly can be avoided, and considering the weighted probabilities of fission trajectories to take the adiabatic fluctuations into account, which can be achieved by using the WKB approximation or Markov chain Monte Carlo sampling based on the
PES.
With these improvements, it is promising to provide systematic studies on nuclear fission and to build a global relationship between fission dynamics
and fragment distribution, which will be helpful in unveiling the mystery of nuclear fission.

\begin{acknowledgement}
		We thank Zhen-Ji Wu and Liang Li for helpful discussions.
This work has been supported by the National Natural Science Foundation of China (Grants No. 12375127 and No. 11975237),
the Strategic Priority Research Program of Chinese Academy of Sciences under Grant No. XDB34010000, and the Fundamental Research Funds for the Central Universities under Grant No. E2E46302.
The results described in this work are obtained on the C3S2 computing center of Huzhou University.
Xiang-Xiang Sun is supported in part by 
NSFC under Grants No. 12205308, and the Deutsche Forschungsgemeinschaft
(DFG) and NSFC through the  funds provided to the Sino-German Collaborative Research Center TRR110  ``Symmetries and the Emergence of  Structure in QCD''
(NSFC Grant No. 12070131001, DFG Project-ID 196253076).
\end{acknowledgement}

%
\bibliographystyle{spphys}
\bibliography{ref}
		
	\end{document}